\documentclass[preprint2]{aastex}

\newcommand{\mearth}{M$_{\oplus}$}
\newcommand{\hh}{H$_2$}
 
\newcommand{\cotwo}{CO$_2$}

\slugcomment{Submitted to the Astrophysical Journal Letters}

\shorttitle{Hydrogen Greenhouse Planets}
\shortauthors{Pierrehumbert \& Gaidos}

\begin{document}


\title{Hydrogen Greenhouse Planets beyond the Habitable Zone}


\author{Raymond Pierrehumbert} \affil{Department of the Geophysical
  Sciences, University of Chicago, Chicago, IL 60637}
\email{rtp1@geosci.uchicago.edu}

\and

\author{Eric Gaidos} \affil{Department of Geology and Geophysics,
  University of Hawaii at Manoa, Honoulu, HI 96822}
\email{gaidos@hawaii.edu}



\begin{abstract}
  We show that collision-induced absorption allows molecular hydrogen to act as
  an incondensible greenhouse gas, and that bars or tens of bars of primordial
  \hh-He mixtures can maintain surface temperatures above the freezing point of
  water well beyond the ``classical'' habitable zone defined for
  \cotwo~greenhouse atmospheres.  Using a 1-D radiative-convective model we find
  that 40 bars of pure $\mathrm{H_2}$ on a 3 Earth-mass planet can maintain a
  surface temperature of 280K out to 1.5AU from an early-type M dwarf star and
  10 AU from a G type star.  Neglecting the effects of clouds and of gaseous
  absorbers besides $\mathrm{H_2}$, the flux at the surface would be sufficient
  for photosynthesis by cyanobacteria (in the G star case) or anoxygenic
  phototrophs (in the M star case).  We argue that primordial atmospheres of one
  to several hundred bars of \hh-He are possible, and use a model of hydrogen
  escape to show that such atmospheres are likely to persist further than 1.5~AU
  from M stars, and 2~AU from G stars, assuming these planets have protecting
  magnetic fields.  We predict that the microlensing planet OGLE-05-390L could
  have retained a \hh-He atmosphere and be habitable at $\sim$2.6~AU from its
  host M star.
\end{abstract}





\section{Introduction}

In the circumstellar habitable zone (HZ), an Earth-like planet has
surface temperatures permissive of liquid water \citep{Hart1979}.  The
HZ is typically calculated assuming an H$_2$O-CO$_2$ greenhouse
atmosphere.  At its inner edge, elevated water vapor leads to a
runaway greenhouse state; at the outer edge, CO$_2$ condenses onto the
surface.  The HZ of the Sun is presently 0.95-2~AU but is much more
compact around less-luminous M dwarfs \citep{Kasting1993}.  Most
reported Earth-size exoplanets are well starward of the HZ because the
Doppler and transit detection techniques are biased towards planets on
close-in orbits \citep{Gaidos2007}.  The {\it Kepler} mission can
detect Earth-size planets in the HZ of solar-type stars
\citep{Koch,Borucki} but these cannot be confirmed by Doppler.  In
contrast, the microlensing technique, in which a planet around a
foreground gravitational lensing star breaks the circular symmetry of
the geometry and produces a transient peak in the light curve of an
amplified background star, is sensitive to planets with projected
separations near the Einstein radius of their host star,
e.g. $\sim$3.5~AU for a solar mass.  Surveys have found planets as
small as 3\mearth~\citep{Bennett2008} and projected separations of
0.7-5~AU \citep{Sumi2010}.  Most microlensing-detected planets orbit M
dwarf stars as these dominate the stellar population.  The Einstein
radius of these stars is well outside the classical HZ
\citep{Sumi2010,Kubas2010} but coincides with the ``ice line'' where
Neptune-mass or icy super-Earth planets may form
\citep{Gould2008,Mann2010}.

The effective temperatures of microlensing planets around M stars will
be below the condensation temperature of all gases except for \hh~and
He.  These mono- and di-atomic gases lack the bending and rotational
modes that impart features to the infrared absorption spectra of more
complex molecules.  At high pressure, however, collisions cause
\hh~molecules to possess transient dipoles moments and induced
absorption.  The resulting greenhouse effect can considerably warm the
surface.  \citet{Stevenson1999} conjectured that `free-floating''
planets with thick \hh-He atmospheres might maintain clement
conditions without benefit of stellar radiation.  Around a star, an
\hh-He atmosphere might admit visible light to the surface: these
planets could, in principle, host photosynthetic life.

We demonstrate that \hh~and He alone can maintain surface temperatures
above the freezing point of water on Earth- to super-Earth-mass
planets beyond the HZ.  In \S \ref{section:RCMethods} we describe
calculations of a radiative-convective atmosphere and its transmission
of light; results are given in \S \ref{section:RCResults}.  In \S
\ref{section:origin} we estimate the amount of \hh~gas that a planet
may accrete from its nascent disk, and the amount lost back to space.
Finally, we discuss the effects of other gases and a candidate
\hh~greenhouse planet detected by microlensing
(\S\ref{section:Discussion}).

\section{Calculation of radiative fluxes and surface temperature}
\label{section:RCMethods}

Surface temperatures were computed using a one-dimensional
radiative-convective model.  For pure $\mathrm{H_2}$ and
$\mathrm{H_2}$-He atmospheres, the infrared opacity due to
collision-induced absorption is from \citet{Borysow2002}.  (Effects of
additional greenhouse gases are discussed in \S
\ref{section:Discussion}.) Thermal infrared fluxes for any given
temperature profile were computed using the integral form of the
solution to the non-scattering Schwarzschild equations, while the
planetary albedo and absorption profile of incoming stellar radiation
were computed using a two-stream scattering code including the effects
of continuum absorption and Rayleigh scattering. The incoming stellar
flux was modeled as blackbodies at 3500 and 6000~K for M and G stars,
respectively. The calculations were independently performed in
wavenumber bands of width 10~cm$^{-1}$ over 10-16490~cm$^{-1}$.  The
atmosphere was assumed not to absorb at higher wavenumbers ($\lambda <
606$~nm), although Rayleigh scattering of incoming stellar radiation
was still considered.  Continuum absorption is a smooth function of
wavenumber, obviating the need for a statistical model within each
band.  Convection was modeled by adjustment to the adiabat appropriate
for a dry $\mathrm{H_2}$-He composition (but see \S
\ref{section:Discussion}).  The adiabat was computed using the ideal
gas equation of state with constant specific heat.


In selected cases up to 10 bar surface pressure $p_s$, the full
radiative-convective model was time-stepped to equilibrium.  The
atmosphere was found to consist of a deep convective troposphere
capped by a nearly isothermal stratosphere. Absorption of incoming
stellar radiation slightly warms the stratosphere, but does not create
an Earth-like temperature inversion because the pressure-dependence of
absorption means that most absorption of stellar radiation and heating
occurs at low altitude or at the surface.  The tropopause occurs at a
pressure level of $\le$170~mbar, producing an optically thin
stratosphere that has little effect on infrared cooling to space.  For
$p_s = 1$~bar, the formation of the stratosphere cools the surface by
only 3~K relative to a calculation in which the entire temperature
profile is adiabatic.  The slight effect becomes even less
consequential as $p_s$ and the infrared optical thickness of the
atmosphere increases (\cite{Pierrehumbert2010}, Chapter 4).
Therefore, we used the computationally efficient "all troposphere"
approximation, in which the temperature profile is set to the adiabat
corresponding to an assumed surface temperature $T_g$.

The radiation code calculates the corresponding infrared flux to space
$OLR(T_g,p_s)$ (Outgoing Longwave Radiation), and planetary albedo
$\alpha(T_g,p_s)$. $T_g$ is obtained from the balance
\begin{equation}
OLR(T_g,p_s) = \frac{1}{4}(1-\alpha(T_g,p_s)) L
\end{equation}
where $L$ is the stellar constant evaluated at the planet's orbit (the
analogue of Earth's Solar constant $L_\odot$). 

\section{Climate model results}
\label{section:RCResults}

\subsection{Minimum surface pressure to maintain surface liquid water}

\begin{figure}
\epsscale{1.0}
\plottwo{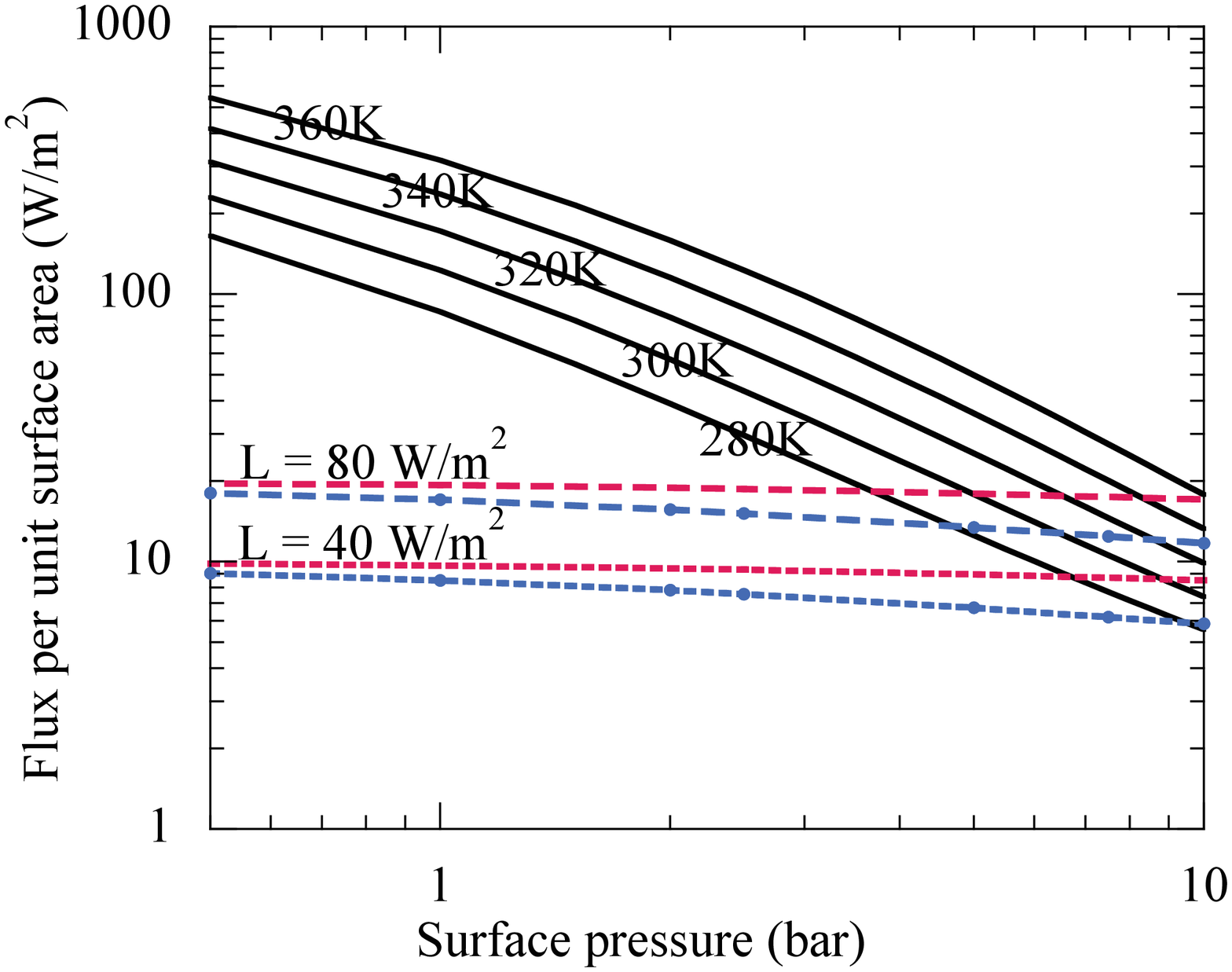}{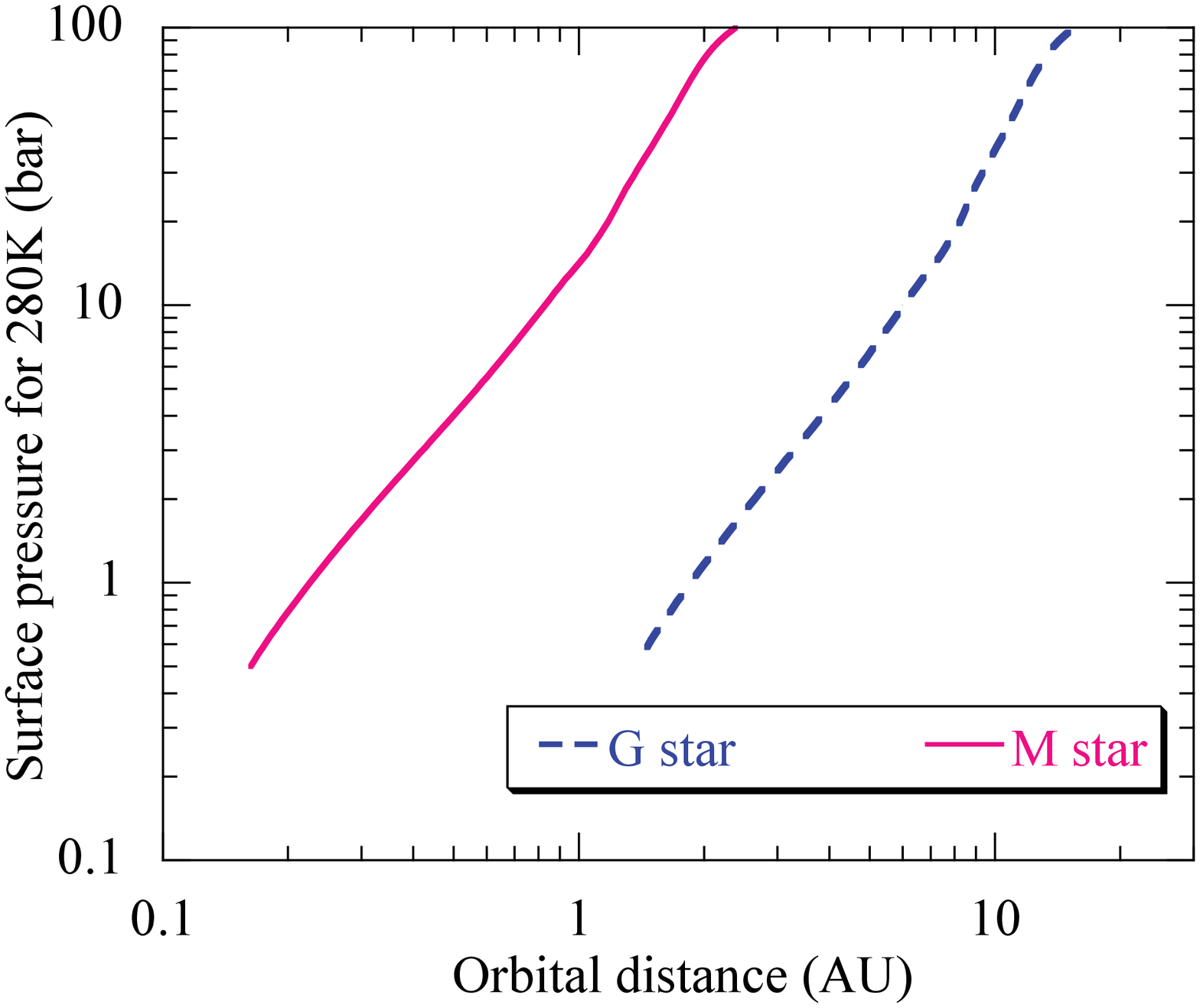}
\caption{Left panel: Determination of surface temperature from the
  top-of-atmosphere radiation budget. Solid curves show the $OLR$
  (infrared emission to space) as a function of surface pressure for
  the various surface temperatures indicated on the curves. Pairs of
  dashed lines give the absorbed solar radiation for stellar constant
  40 $\mathrm{W m^{-2}}$ (short dashes) and 80 $\mathrm{W m^{-2}}$
  (long dashes).  The upper curve in each pair is for an M star
  spectrum while the lower is for a G star.  Right panel: Surface
  pressure required to maintain 280K surface temperature, as a
  function of radius of a circular orbit.  Results are given for an M
  star (0.013 times solar luminosity), and a G star (solar
  luminosity).  All calculations were carried out for a pure
  $\mathrm{H_2}$ atmosphere on a planet with surface gravity 17
  $\mathrm{m s^{-2}}$ Planetary albedos were computed assuming zero
  surface albedo. \label{fig.Climate}}
\end{figure}

We present results for a 3\mearth~rocky planet with surface gravity $g
= 17$ m s$^{-2}$ \citep{Seager2007}. Gravity enters the calculation of
$OLR$ only in the combination $p_s^2/g$; the results can be applied
approximately to other values of $g$ by scaling $p_s$. Scaled results
will diverge slightly from the correct values because Rayleigh
scattering depends on $p_s/g$, whence albedo scales differently from
$OLR$.  Figure \ref{fig.Climate} (left panel) illustrates the energy
balance for a pure $\mathrm{H_2}$ atmosphere.  $OLR$ computed with the
all-troposphere approximation is shown as a function of $p_s$ for
different values of $T_g$, together with curves of absorbed stellar
radiation for G and M star cases with stellar constants $L = 40$ W
m$^{-2}$ and $L = 80$ W m$^{-2}$.  Equilibria are represented by the
intersections between $OLR$ curves and absorbed stellar radiation
curves.  The absorbed stellar radiation decreases more rapidly with
$p_s$ for G stars than for M stars because the greater shortwave
radiation in the former case leads to more Rayleigh scattering,
whereas the latter is dominated by absorption. For $L = 40$ W m$^{-2}$
(somewhat less than Jupiter's insolation) $p_s = 10$~bar is sufficient
to maintain a surface temperature of 280~K about a G star; around an M
star, only 7 bars is required.

Figure \ref{fig.Climate} (right panel) also shows the minimum $p_s$
required to maintain $T_g = 280$~K as a function of the orbital
distance.  We use 280~K as a criterion because an ocean world whose
global mean temperature is too near freezing is likely to enter a
``snowball'' state \citep{Pierrehumbert2011}.  The required pressure
is higher around M stars because more greenhouse effect is needed to
compensate for the lower luminosity.  For an early M-type star with
luminosity 1.3\% of the Sun, 100~bars of $\mathrm{H_2}$ maintains
liquid surface water out to 2.4~AU; around a G star, these conditions
persist to 15~AU.  Cases requiring $>20$ bars are speculative because
the maintenance of a deep convective troposphere, when so little
stellar radiation reaches the surface, is sensitive to small
admixtures of strongly shortwave-absorbing constituents.



Our results are not appreciably affected by the formation of a
stratosphere.  The stratosphere in these atmospheres is essentially
transparent to incoming stellar radiation and heated mainly by
absorption of upwelling infrared radiation.  As expected for a gas
with frequency-dependent absorptivity, the stratospheric temperature
is somewhat (few K) below the grey-gas skin temperature
$T_{eff}/2^{1/4}$, where $\sigma T_{eff}^4 = OLR =
\frac{1}{4}(1-\alpha)L$ (\cite{Pierrehumbert2010}, Chapter 4).  


\subsection{Photosynthetic Active Radiation at the
  Surface \label{section:par}}



\begin{figure}
\epsscale{1.0}
\plotone{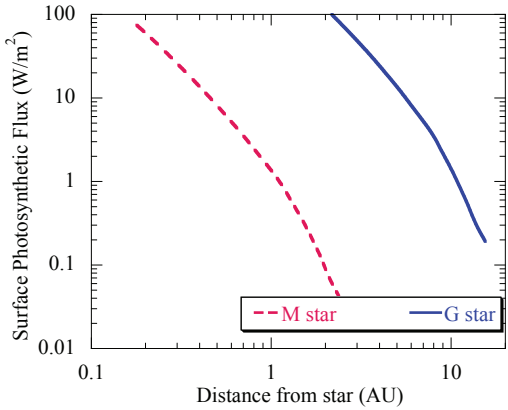}
\caption{Flux in the photosynthetically active band (400-700nm)
  reaching the surface at the substellar point as a function of
  distance from a G or M host star. The flux is calculated assuming an
  $\mathrm{H_2}$ atmosphere with surface pressure sufficient to
  maintain 280~K surface temperature at each
  distance. \label{fig:PhotoFlux}}
\end{figure}

Starlight is a potential energy source for life, so we consider
whether the stellar flux reaching the surface is sufficient to sustain
photosynthesis.  Most terrestrial photosynthetic organisms use light
in the $\lambda=400-700$~nm range.  The flux to sustain half maximum
growth rate for many cyanobacteria is $\sim$1~W m$^{-2}$
\citep{Carr1982,Tilzer1987}, and an anoxygenic green sulfur bacterium
can use a flux of $3 \times 10^{-3}$~$\mathrm{W m^{-2}}$
\citep{Manske2005}.  Planets on more distant orbits experience less
stellar irradiation and require a thicker atmosphere to maintain
surface liquid water, which reduce transmission of light to the
surface.  For a pure $\mathrm{H_2}$ atmosphere, transmission in
photosynthetically active radiation (PAR) is limited by Rayleigh
scattering and differs slightly for M vs. G star spectra. The main
difference is that G stars have higher luminosity, requiring less
atmosphere to maintain liquid water at a given orbital distance, and a
higher proportion of their output is PAR.  Figure \ref{fig:PhotoFlux}
shows surface PAR at the substellar point as a function of orbital
distance for a minimum pure \hh~atmosphere to maintain 280~K.
Hydrogen greenhouse planets as far as 10~AU from a G star or 1~AU from
an M star could sustain cyanobacteria-like life; distances of tens of
AU or several AU, respectively, still permit organisms like anoxygenic
phototrophs.

\section{Origin and Loss of H-He Atmospheres \label{section:origin}}

A body whose Bondi radius (at which gravitational potential energy
equals gas enthalpy) exceeds its physical radius will accumulate gas
from any surrounding disk.  The critical mass is less than a lunar
mass and protoplanets may acquire primordial atmospheres of \hh~and
He, as well as gases released by impacts \citep{Stevenson}.
Earth-mass planets will not experience runaway gas accretion and
transformation into gas giants \citep{Mizuno1980}.  The atmospheres of
accreting Earth-mass planets will probably be optically thick and
possess an outer radiative zone \citep{Rafikov2006}.  If the gas
opacity is supplied by heavy elements and is pressure-independent, the
atmosphere mass is inversely proportional to the accretion luminosity
\citep{Stevenson,Ikoma2006,Rafikov2006}.  In the ``high accretion
rate'' case (planetesimals are strongly damped by gas drag), the
surface pressure will be $1.2a^{1.5}$($M_p$/\mearth) bar, where $a$ is
the orbital semimajor axis in AU and $M_p$ is the planet mass.  In the
``intermediate accretion rate'' case (gas drag is unimportant), the
surface pressure will be $16a^{2}$($M_p$/\mearth)~bar
\citep{Rafikov2006}.  At 3~AU, the predicted timescales for these two
phases of accretion are $\sim10^6$ and $10^7$~yr, respectively, and
bracket the lifetime of gaseous disks \citep{HaischJr.2001,Evans2009}.
Thus, primordial atmospheres with surface pressures of one to several
hundred bars are plausible.  Hydrogen is also released during the
reaction of water with metallic iron \citep{ElkinsTanton2008}.

Giant impacts may erode the proto-atmosphere, but not completely
remove it \citep{Genda2003,Genda2004}, although the mechanical
properties of an ocean may amplify such effects \citep{Genda2005}.
Instead, the low molecular weight $\mu$ of atomic H makes it
susceptible to escape to space when stellar extreme ultraviolet (EUV)
radiation heats the upper atmosphere, and by stellar winds. We examine
EUV-driven escape of atomic H from a planet of radius $r_p$ using a
thermal model of an H- or \hh-dominated upper atmosphere, assuming the
escape rate is not limited by \hh~dissociation.  The upper boundary is
at the exobase where H escapes, and the lower boundary is where eddy diffusivity
becomes important (in essence, the homopause). 
At the homopause, the number density is taken to be
$n = 10^{19}$~m$^{-3}$ \citep{Yamanaka1995}, and the temperature is set to the stratosphere temperatures predicted by the
radiative-convective model (see \S \ref{section:RCResults}).
The exobase occurs at an altitude
$h_e$ where the cross-section for collisions with molecular hydrogen
is $\sim1$, i.e. at a column density $N = 2.3 \times
10^{19}$~m$^{-2}$. The exobase temperature $T_e$ and Jeans parameter
$\lambda = GM_p\mu/((r_p + h_e)kT_e)$ are calculated by balancing the
globally-averaged EUV heating $q_{EUV}$ with heat lost to conduction,
radiative cooling, and escape.  At $\lambda > 2.8$ we use the Jeans
escape rate (in surface pressure per unit time):
\begin{equation}
  \phi = \frac{(GM_p)^2\mu}{\left(r_p+h_e\right)^4}\sqrt{\frac{\mu}{2\pi kT_e}}N_e \left(1+\lambda\right)e^{-\lambda},
\end{equation}
where $N_e$ is the number density at the exobase.  For Jeans escape, we assume H is the
escape component and that H$_2$ is efficiently dissociated below the
exobase; this maximizes the escape rate.  When $\lambda \le 2.8$,
marking the transition to hydrodynamic escape \citep{Volkov2010}, we
assume that escape is energy-limited:
\begin{equation}
\label{eqn.energylimited}
  \phi = q_{EUV} \frac{\left(r_p + h_t\right)^2}{r_p^3},
\end{equation}
where $h_t$ is the altitude where EUV is absorbed (the thermosphere),
and the right-hand factor accounts for thermal expansion.

All EUV ($\lambda < 912$\AA) is assumed to be absorbed in a
layer at $N = 2.3 \times 10^{19}$~m$^{-2}$.  Lyman-$\alpha$ radiation
resonantly scatters at higher altitudes ($N \sim1 \times
10^{17}$~m$^{-2}$) but, in the absence of an absorber such as atomic
oxygen, will escape before being thermalized
\citep{Murray-Clay2009}. 

For given exobase and homopause temperatures, the temperature profile
is calculated downward to its maximum at the thermosphere and then to
the homopause, assuming a mean $\mu$ of 3.1 (appropriate
for \hh+He in a solar proportion).  The exobase temperature is
adjusted until the correct value of $n$ at the homopause is reached.
Cooling from the escape of H at the exobase, the adiabatic upward
motion of gas in the atmosphere to maintain steady-state, and the
collision-induced opacity of \hh~ \citep{Borysow1997,Borysow2002} are
included.  We use a thermal conductivity $\mathrm{k \approx
  0.0027~T^{0.73}~W~m^{-1}~K^{-1}}$ based on \citet{Allison1971}.  We
adopt $q_{EUV} = 1.5 \times 10^{-4}$~W~m$^{-2}$ at 1~AU from a
4.5~Gyr-old G star \citep{Watson1981}.  This is the current solar
incident EUV flux of $1.67 \times 10^{-3}$~W~m$^{-2}$ for $\lambda <
900$\AA~\citep{Hinteregger1981,DelZanna2010} times a 30\% heating
efficiency \citep{Watson1981}.  
EUV is scaled to different ages and M stars
assuming proportionality to soft X-ray flux \citep{Penz2008, Penz2008a}.

At sufficiently high EUV flux, the atmosphere expands to the scale of
the planet, gravity at the exobase decreases, no steady-state solution to the
temperature profile exists, and $\lambda < 2.8$.  We presume this to
correspond to hydrodynamic escape, fix the temperature structure at
its last state consistent with Jeans escape, and use the
energy-limited escape rate (Equation \ref{eqn.energylimited}).
Additional H may escape as a result of interaction with the stellar
wind \citep{Lammer2007}.

We calculate total atmospheric lost by 4.5~Gyr for planets around G
and M dwarf stars (Figure \ref{fig.loss}).  Significant atmospheric
loss usually ceases by $\sim$2~Gyr.  Planets around M dwarfs, with
their lower X-ray luminosities, can retain their \hh-He atmospheres
close to the star.  However, these stars, especially those of mid- to
late-M spectral type, may vary greatly in their EUV flux
\citep{Reiners2008,Browning2010}.

\section{Discussion \label{section:Discussion}}

{\it Effects of other gases:} The addition of He to the pure $\mathrm{H_2}$
atmospheres considered above would warm the surface by somewhat less than the
effect of adding the same number of $\mathrm{H_2}$ molecules, because the CIA
coefficients for $\mathrm{H_2}$-He collision are generally weaker than those for
$\mathrm{H_2}$-$\mathrm{H_2}$ \citep{Borysow1988}, and because He-He collisions do not
contribute. Addition of He also makes the adiabat steeper and the surface
warmer relative to the pure $\mathrm{H_2}$ case, but for a 10\% He concentration
the extra warming is no more than 1~K.

The warming provided by the $\mathrm{H_2}$ greenhouse permits other
greenhouse gases to be stable against complete condensation, notably
$\mathrm{CO_2}$, $\mathrm{H_2O}$, $\mathrm{NH_3}$ and $\mathrm{CH_4}$.
For the distant orbits of interest in this Letter, the effective
radiating temperature occurs in the upper troposphere, and is
$\sim$60~K for a planet 2~AU from an M star.  The low saturation vapor
pressure at this temperature confines the gases to lower altitudes,
where they add little to the greenhouse effect.  Even at 100~K the
saturation vapor pressures of $\mathrm{CO_2}$, $\mathrm{H_2O}$,
$\mathrm{NH_3}$ are all $<$0.02~Pa.  $\mathrm{CH_4}$, with a vapor
pressure of 0.338 bar at 100~K, is most likely to exert a modest
additional warming effect.

The condensible gases also affect climate through latent heat release,
which makes the adiabat less steep and reduces the surface
temperature. Because of the heat capacity of the thick $\mathrm{H_2}$
atmosphere, the effect on the adiabat is slight, amounting to, for
example, a 3~K surface cooling for an $\mathrm{H_2O}$ saturated
atmosphere in 5 bars of $\mathrm{H_2}$.  It is much less for thicker
atmospheres.  Instead, the main effect of condensible constituents
would be to form tropospheric clouds, which would increase the albedo
and hence increase the $\mathrm{H_2}$ pressure needed to maintain
habitability. A quantitative treatment of clouds is outside the scope
of this Letter.

\begin{figure}
\epsscale{1}
\plotone{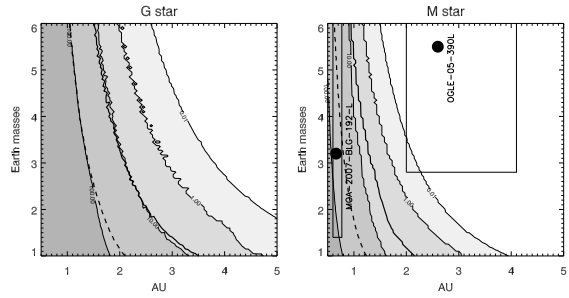}
\caption{EUV-driven atmospheric escape (bars) at an age of 4.5~Gry as
  a function of planet mass and distance from a G star (left) or M
  star (right).  The heavy solid line is where mass loss equals the
  estimated mass of proto-atmosphere acquired by a planet during its
  early phase of oligarchic growth (high rate of planetesimal
  accretion) and the heavy dashed line is the same for the case of the
  late phase of oligarchic growth (intermediate rate of planetesimal
  accretion).  See text for details.  Two microlensing-detected
  super-Earth-mass planets are plotted, with the boxes representing
  the undercertainties in their parameters.  \label{fig.loss}}
\end{figure}

{\it Cold cases:} Two ``super-Earth''-mass planets detected by
microlensing are plotted in Figure \ref{fig.loss}b.  MOA-2007-BLG-192L
orbits $\sim$0.7~AU from a very late-type M dwarf.  It has an
effective temperature of 40-50~K, below the condensation temperature
of all gases except \hh~and He, although N$_2$ or CO might be
volatilized at the substellar point of a synchronously rotating planet
\citep{Kubas2010}.  We predict that any primordial hydrogen atmosphere has been
lost by EUV-driven hydrodynamic escape carrying
He and other light volatiles with it.  
In contrast, more massive OGLE-05-390L, which orbits
$\sim$2.6~AU from a mid M-type star but has a similar effective
temperature \citep{Ehrenreich2006}, may retain a primordial hydrogen
atmosphere.  This planet could potentially sustain liquid water at its
surface, and may represent potentially ocean-bearing
planets to be revealed by future microlensing surveys
\citep{Beaulieu2008,Bennett2010}.

All planets smaller than Neptune detected by {\it Kepler} or
Doppler orbit within the zone where hydrogen escape is predicted to be
efficient.  Gliese~581d ($\ge 6$\mearth, 0.2~AU) is in this category;
though it may lie within the classic HZ
\citep{Wordsworth2010,vonParis2010,Hu2011}.  Larger, transiting
planets such as GJ~436b (22~\mearth) have retained a low
molecular-weight envelope despite their proximity to their parent
stars, perhaps due to their high gravity and migration from further
out in the primordial nebula.  Should the {\it Kepler} mission be extended
($\sim$6~yr), planets at 1.5-2~AU, the inner boundary where a hydrogen atmosphere
is retained (Figure \ref{fig.loss}), could be confirmed with 3 transits.  
Measuring mass would be at the limit of current Doppler technology
(radial velocity amplitude $\le$ 1~m~s$^{-1}$), but spectra obtained during 
transits might reveal any H$_2$-rich atmosphere.   
 
{\it The effects of life:} These habitable worlds may nurture the
seeds of their own destruction, on account of the chemical
disequilibrium between an \hh-dominated atmosphere and (we presume) a
comparatively oxidizing silicate mantle. If the planet's tectonics
supports $\mathrm{CO_2}$ outgassing as is the case for Earth, then
methanogens could deplete the $\mathrm{H_2}$ atmosphere by combining
it with $\mathrm{CO_2}$ to produce $\mathrm{CH_4}$.  The
$\mathrm{O_2}$ produced by cyanobacteria would similarly consume
$\mathrm{H_2}$ and convert the atmosphere to organic carbon and water. Either case
would eventually cause a freeze-out of the atmosphere, in the absence
of some biotic or abiotic process that regenerates $\mathrm{H_2}$.

\acknowledgments

This research was supported by NASA grant NNX10AQ36G and was inspired
by discussions between EG, RP, Scott Gaudi, and Sara Seager at the
Aspen Institute for Physics.  We acknowledge two anonymous reviewers
of this manuscript.


\begin{thebibliography}{50}
\expandafter\ifx\csname natexlab\endcsname\relax\def\natexlab#1{#1}\fi

\bibitem[{Allison \& Smith(1971)}]{Allison1971}
Allison, A.~C., \& Smith, F.~J. 1971, Atomic Data and Nuclear Data Tables, 3,
  371

\bibitem[{Beaulieu {et~al.}(2008)Beaulieu, Kerins, Mao, \&
  Bennett}]{Beaulieu2008}
Beaulieu, J., Kerins, E., Mao, S., \& Bennett, D. 2008, Arxiv preprint arXiv:,
  0808.0005B

\bibitem[{Bennett {et~al.}(2010)Bennett, Anderson, Beaulieu, Bond, Cheng, Cook,
  Friedman, Gaudi, Gould, Jenkins, \& Others}]{Bennett2010}
Bennett, D., {et~al.} 2010, Arxiv preprint arXiv:1012.4486

\bibitem[{Bennett {et~al.}(2008)Bennett, Bond, Udalski, Sumi, Abe, Fukui,
  Furusawa, Hearnshaw, Holderness, Itow, Kamiya, Korpela, Kilmartin, Lin, Ling,
  Masuda, Matsubara, Miyake, Muraki, Nagaya, Okumura, Ohnishi, Perrott,
  Rattenbury, Sako, Saito, Sato, Skuljan, Sullivan, Sweatman, Tristram, Yock,
  Kubiak, Szymański, Pietrzyński, Soszyński, Szewczyk, Wyrzykowski, Ulaczyk,
  Batista, Beaulieu, Brillant, Cassan, Fouqu\'{e}, Kervella, Kubas, \&
  Marquette}]{Bennett2008}
Bennett, D.~P., {et~al.} 2008, The Astrophysical Journal, 684, 663

\bibitem[{Borucki {et~al.}(2011)Borucki, Koch, Basri, Batalha, Brown, Bryson,
  Caldwell, Christensen-dalsgaard, Cochran, Devore, Dunham, Iii, Geary, Gould,
  Howell, Jenkins, Latham, J, Ciardi, Doyle, Dupree, Ford, Holman, Seager,
  Steffen, Tarter, F, D\'{e}sert, Endl, Fabrycky, Fressin, \& Haas}]{Borucki}
Borucki, W.~J., {et~al.} 2011, Arxiv preprint arXiv:1102.0541, 1

\bibitem[{Borysow(2002)}]{Borysow2002}
Borysow, A. 2002, Astronomy and Astrophysics, 390, 779

\bibitem[{Borysow {et~al.}(1988)Borysow, Frommhold, \& Birnbaum}]{Borysow1988}
Borysow, A., Frommhold, L., \& Birnbaum. 1988, The Astrophysical journal, 329,
  509

\bibitem[{Borysow {et~al.}(1997)Borysow, Jorgensen, \& Zheng}]{Borysow1997}
Borysow, A., Jorgensen, U., \& Zheng, C. 1997, Astronomy and Astrophysics, 324,
  185

\bibitem[{Browning {et~al.}(2010)Browning, Basri, Marcy, West, \&
  Zhang}]{Browning2010}
Browning, M.~K., Basri, G., Marcy, G.~W., West, A.~A., \& Zhang, J. 2010,
  Astronomical Journal, 139, 504

\bibitem[{Carr \& Whitton(1982)}]{Carr1982}
Carr, N.~G., \& Whitton, B.~A. 1982, {The Biology of Cyanobacteria} (Berkeley:
  University of California Press), 688

\bibitem[{{Del Zanna} {et~al.}(2010){Del Zanna}, Andretta, Chamberlin, Woods,
  \& Thompson}]{DelZanna2010}
{Del Zanna}, G., Andretta, V., Chamberlin, P., Woods, T., \& Thompson, W. 2010,
  Astronomy and Astrophysics, 518, A49

\bibitem[{Ehrenreich {et~al.}(2006)Ehrenreich, {Lecavelier des Etangs},
  Beaulieu, \& Grasset}]{Ehrenreich2006}
Ehrenreich, D., {Lecavelier des Etangs}, A., Beaulieu, J., \& Grasset, O. 2006,
  The Astrophysical Journal, 651, 535

\bibitem[{Elkins‐Tanton \& Seager(2008)}]{ElkinsTanton2008}
Elkins‐Tanton, L.~T., \& Seager, S. 2008, The Astrophysical Journal, 685,
  1237

\bibitem[{Evans {et~al.}(2009)Evans, Dunham, J\o~rgensen, Enoch, Mer\'{\i}n,
  van Dishoeck, Alcal\'{a}, Myers, Stapelfeldt, Huard, Allen, Harvey, van
  Kempen, Blake, Koerner, Mundy, Padgett, \& Sargent}]{Evans2009}
Evans, N.~J., {et~al.} 2009, The Astrophysical Journal Supplement Series, 181,
  321

\bibitem[{Gaidos {et~al.}(2007)Gaidos, Haghighipour, Agol, Latham, Raymond, \&
  Rayner}]{Gaidos2007}
Gaidos, E., Haghighipour, N., Agol, E., Latham, D., Raymond, S.~N., \& Rayner,
  J. 2007, Science, 318, 210

\bibitem[{Genda \& Abe(2003)}]{Genda2003}
Genda, H., \& Abe, Y. 2003, Icarus, 164, 149

\bibitem[{Genda \& Abe(2004)}]{Genda2004}
---. 2004, Lunar Planet. Sci, 1518

\bibitem[{Genda \& Abe(2005)}]{Genda2005}
---. 2005, Nature, 433, 842

\bibitem[{Gould {et~al.}(2010)Gould, Dong, Gaudi, Udalski, Bond, Greenhill,
  Street, Dominik, Sumi, Szyma$\backslash$'nski, \& Others}]{Gould2008}
Gould, A., {et~al.} 2010, The Astrophysical Journal, 720, 1073

\bibitem[{{Haisch Jr.} {et~al.}(2001){Haisch Jr.}, Lada, \&
  Lada}]{HaischJr.2001}
{Haisch Jr.}, K.~E., Lada, E.~A., \& Lada, C.~J. 2001, Astrophysical Journal,
  553, L153

\bibitem[{Hart(1979)}]{Hart1979}
Hart, M.~H. 1979, Icarus, 37, 351

\bibitem[{Hinteregger {et~al.}(1981)Hinteregger, Fukui, \&
  Gilson}]{Hinteregger1981}
Hinteregger, H.~E., Fukui, K., \& Gilson, B.~R. 1981, Geophysical Research
  Letters, 8, 1147

\bibitem[{Hu \& Ding(2011)}]{Hu2011}
Hu, Y., \& Ding, F. 2011, Astronomy and Astrophysics, 526, A135

\bibitem[{Ikoma \& Genda(2006)}]{Ikoma2006}
Ikoma, M., \& Genda, H. 2006, The Astrophysical Journal, 648, 696

\bibitem[{Kasting {et~al.}(1993)Kasting, Whitmire, \& Reynolds}]{Kasting1993}
Kasting, J.~F., Whitmire, D.~P., \& Reynolds, R.~T. 1993, Icarus, 101, 108

\bibitem[{Koch {et~al.}(2010)Koch, Borucki, Basri, Batalha, Brown, Caldwell,
  Christensen-Dalsgaard, Cochran, DeVore, Dunham, \& Others}]{Koch}
Koch, D., {et~al.} 2010, The Astrophysical Journal Letters, 713, L79

\bibitem[{Kubas {et~al.}(2010)Kubas, Beaulieu, Bennett, Cassan, Cole, Lunine,
  Marquette, Dong, Gould, Sumi, \& Others}]{Kubas2010}
Kubas, D., {et~al.} 2010, Arxiv preprint arXiv:1009.5665

\bibitem[{Lammer {et~al.}(2007)Lammer, Lichtenegger, Kulikov, Griessmeier,
  Terada, Erkaev, Biernat, Khodachenko, RIbas, Penz, \& Selsis}]{Lammer2007}
Lammer, H., {et~al.} 2007, Astrobiology, 7, 185

\bibitem[{Leger {et~al.}(2004)Leger, Selsis, Sotin, Guillot, Despois, Lammer,
  Ollivier, Brachet, Labeque, \& Valette}]{Leger2004a}
Leger, A., {et~al.} 2004, Icarus, 169, 499

\bibitem[{Mann {et~al.}(2010)Mann, Gaidos, \& Gaudi}]{Mann2010}
Mann, A., Gaidos, E., \& Gaudi, B. 2010, The Astrophysical Journal, 719, 1454

\bibitem[{Manske {et~al.}(2005)Manske, Glaeser, Kuypers, \&
  Overmann}]{Manske2005}
Manske, A., Glaeser, J., Kuypers, M., \& Overmann, J. 2005, Applied and
  environmental microbiology, 71, 8049

\bibitem[{Mizuno(1980)}]{Mizuno1980}
Mizuno, H. 1980, Progress in Theoretical Physics, 64, 544

\bibitem[{Murray-Clay {et~al.}(2009)Murray-Clay, Chiang, \&
  Murray}]{Murray-Clay2009}
Murray-Clay, R.~a., Chiang, E.~I., \& Murray, N. 2009, The Astrophysical
  Journal, 693, 23

\bibitem[{Penz \& Micela(2008)}]{Penz2008a}
Penz, T., \& Micela, G. 2008, Astronomy, 584, 579

\bibitem[{Penz {et~al.}(2008)Penz, Micela, \& Lammer}]{Penz2008}
Penz, T., Micela, G., \& Lammer, H. 2008, Astronomy, 314, 309

\bibitem[{Pierrehumbert(2010)}]{Pierrehumbert2010}
Pierrehumbert, R.~T. 2010, {Principles of Planetary Climate} (Cambridge:
  Cambridge University Press)

\bibitem[{Pierrehumbert {et~al.}(2011)Pierrehumbert, Abbot, Voigt, \&
  Koll}]{Pierrehumbert2011}
Pierrehumbert, R.~T., Abbot, D., Voigt, A., \& Koll, D. 2011, Annual Review of
  Earth and Planetary Science, 39, DOI:10.1146

\bibitem[{Rafikov(2006)}]{Rafikov2006}
Rafikov, R. 2006, The Astrophysical Journal, 648, 666

\bibitem[{Reiners \& Basri(2008)}]{Reiners2008}
Reiners, A., \& Basri, G. 2008, Astrophysical Journal, 684, 1390

\bibitem[{Seager {et~al.}(2007)Seager, Kuchner, Hier‐Majumder, \&
  Militzer}]{Seager2007}
Seager, S., Kuchner, M., Hier‐Majumder, C.~a., \& Militzer, B. 2007, The
  Astrophysical Journal, 669, 1279

\bibitem[{Stevenson(1982)}]{Stevenson}
Stevenson, D.~J. 1982, Planetary and Space Science, 30, 755

\bibitem[{Stevenson(1999)}]{Stevenson1999}
---. 1999, Nature, 400, 32

\bibitem[{Sumi {et~al.}(2010)Sumi, Bennett, Bond, Udalski, Batista, Dominik,
  Fouqu\'{e}, Kubas, Gould, Macintosh, Cook, Dong, Skuljan, Cassan, Abe,
  Botzler, Fukui, Furusawa, Hearnshaw, Itow, Kamiya, Kilmartin, Korpela, Lin,
  Ling, Masuda, Matsubara, Miyake, Muraki, Nagaya, Nagayama, Ohnishi, Okumura,
  Perrott, Rattenbury, Saito, Sako, Sullivan, Sweatman, Tristram, Yock,
  Beaulieu, Cole, Coutures, Duran, Greenhill, Jablonski, Marboeuf, Martioli,
  Pedretti, Pejcha, Rojo, Albrow, Brillant, Bode, Bramich, Burgdorf, Caldwell,
  Calitz, Corrales, Dieters, Prester, Donatowicz, Hill, Hoffman, Horne,
  J\o~rgensen, Kains, Kane, Marquette, Martin, Meintjes, Menzies, Pollard,
  Sahu, Snodgrass, Steele, Street, Tsapras, Wambsganss, Williams, Zub,
  Szymański, Kubiak, Pietrzyński, Soszyński, Szewczyk, Wyrzykowski, Ulaczyk,
  Allen, Christie, DePoy, Gaudi, Han, Janczak, Lee, McCormick, Mallia, Monard,
  Natusch, Park, Pogge, \& Santallo}]{Sumi2010}
Sumi, T., {et~al.} 2010, The Astrophysical Journal, 710, 1641

\bibitem[{Tilzer(1987)}]{Tilzer1987}
Tilzer, M. 1987, New Zealand Journal of Marine and Freshwater Research, 21, 401

\bibitem[{Volkov {et~al.}(2010)Volkov, Johnson, Tucker, \& Erwin}]{Volkov2010}
Volkov, A., Johnson, R., Tucker, O., \& Erwin, J. 2010, Arxiv preprint
  arXiv:1009.5110, 1

\bibitem[{von Paris {et~al.}(2010)von Paris, Gebauer, Godolt, Grenfell, Hedelt,
  Kitzmann, Patzer, Rauer, \& Stracke}]{vonParis2010}
von Paris, P., {et~al.} 2010, Astronomy and Astrophysics, 522, A23

\bibitem[{Watson {et~al.}(1981)Watson, Donahue, \& Walker}]{Watson1981}
Watson, A.~J., Donahue, T.~M., \& Walker, J. C.~G. 1981, Icarus, 48, 150

\bibitem[{Wolf \& Toon(2010)}]{Wolf2010}
Wolf, E., \& Toon, O. 2010, Science, 328, 1266

\bibitem[{Wordsworth {et~al.}(2010)Wordsworth, Forget, Selsis, Madeleine,
  Millour, \& Eymet}]{Wordsworth2010}
Wordsworth, R., Forget, F., Selsis, F., Madeleine, J., Millour, E., \& Eymet,
  V. 2010, Astronomy and Astrophysics, 522, A22

\bibitem[{Yamanaka(1995)}]{Yamanaka1995}
Yamanaka, M. 1995, Advances in Space Research, 15, 47

\end{thebibliography}

\clearpage



\end{document}